\documentclass[10pt,twocolumn,letterpaper]{article}

\usepackage[pagenumbers]{cvpr}
\usepackage[normalem]{ulem}
\useunder{\uline}{\ul}{}
\usepackage[accsupp]{axessibility}  
\definecolor{cvprblue}{rgb}{0.21,0.49,0.74}
\usepackage[pagebackref,breaklinks,colorlinks,allcolors=cvprblue]{hyperref}

\def\paperID{7} 
\def\confName{CVPR}
\def\confYear{2025}

\title{Prototype-Guided Diffusion for Digital Pathology: Achieving Foundation Model Performance with Minimal Clinical Data}

\author{
Ekaterina Redekop\\
{\tt\small eredekop@g.ucla.edu}
\and
Mara Pleasure\\
{\tt\small mpleasure@g.ucla.edu}
\and
Vedrana Ivezic\\
{\tt\small vivezic@g.ucla.edu}
\and
Zichen Wang\\
{\tt\small zcwang0702@ucla.edu}
\and
Kimberly Flores\\
{\tt\small KimberlyFlores@mednet.ucla.edu}
\and
Anthony Sisk\\
{\tt\small ASisk@mednet.ucla.edu}
\and
William Speier\\
{\tt\small Speier@ucla.edu}
\and
Corey Arnold\\
{\tt\small cwarnold@ucla.edu}
\and
\centerline{University of California, Los Angeles, USA}
}

\begin{document}
\maketitle
\begin{abstract}
Foundation models in digital pathology use massive datasets to learn useful compact feature representations of complex histology images. However, there is limited transparency into what drives the correlation between dataset size and performance, raising the question of whether simply adding more data to increase performance is always necessary.
In this study, we propose a prototype-guided diffusion model to generate high-fidelity synthetic pathology data at scale, enabling large-scale self-supervised learning and reducing reliance on real patient samples while preserving downstream performance. Using guidance from histological prototypes during sampling, our approach ensures biologically and diagnostically meaningful variations in the generated data. 
We demonstrate that self-supervised features trained on our synthetic dataset achieve competitive performance despite using $\thicksim$60x–760x less data than models trained on large real-world datasets. Notably, models trained using our synthetic data showed statistically comparable or better performance across multiple evaluation metrics and tasks, even when compared to models trained on orders of magnitude larger datasets. Our hybrid approach, combining synthetic and real data, further enhanced performance, achieving top results in several evaluations. These findings underscore the potential of generative AI to create compelling training data for digital pathology, significantly reducing the reliance on extensive clinical datasets and highlighting the efficiency of our approach.
\end{abstract}    
\section{Introduction}
\label{sec:intro}

Recent advances in deep learning for digital pathology have been powered by large-scale foundation models trained on extensive histopathology datasets with self-supervised learning (SSL). 
This research has primarily focused on scaling up dataset sizes, driven by the assumption that more data inherently improves performance. One of the first examples, UNI, is a foundation model for pathology, pre-trained using more than 100 million images \cite{chen2024towards}. Another widely used recent foundational model, Prov-GigaPath, was pre-trained on 1.3 billion pathology images \cite{xu2024whole}. This trend is further reflected in efforts to expand pathology image repositories based on existing evidence that indicates increasing data volume consistently improves downstream performance \cite{bilal2025foundation}. While this notion is intuitive and supported by experimental results, the exact technical reasons for increased performance have been less thoroughly explored, and more importantly, there is little work showing if alternative methods could produce similar results. One alternative to accumulating more data is to use generative modeling to artificially expand training datasets \cite{xue2021selective, xue2019synthetic, ye2020synthetic}. By utilizing data augmentation to enrich the diversity and quality of pretraining datasets, the downstream performance can be improved. For example, it has been shown that increasing the ratio of synthetic to real data to 100\% increases the classifier’s accuracy by up to 6.4\% and F1 score by up to 6.6\% \cite{ye2020synthetic, alimisis2025advances, zhu2023domainstudio}. In the field of medical imaging, such an approach would be advantageous given the challenges of collecting medical data and ensuring its security and compliant use.
Diffusion models, through their iterative noise-and-denoising learning process, have emerged as a powerful tool for dataset augmentation, surpassing the performance of competing generative models such as generative adversarial networks (GANs) \cite{goodfellow2020generative}. Moreover, conditioning a diffusion model's generation process on various inputs, such as class labels or text, allows for precise control over image creation. This capability to support task-specific augmentation has sparked active research into synthetic data generation for digital pathology \cite{pozzi2024generating, pozzi2023generating, linmans2024diffusion, sidulova2024contextual, alimisis2025advances}. 

\begin{figure*}[h!]
    \centering
    \includegraphics[scale=0.95]{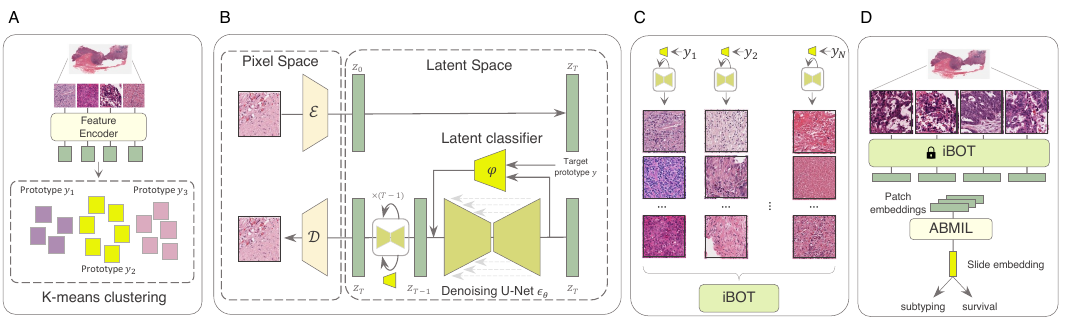} %
    \caption{Overview of the proposed approach. \textbf{A.} A WSI is segmented and patched into a set of non-overlapping patches. A compressed feature for each patch is obtained through a pre-trained feature encoder. K-Means clustering is performed to identify prototypes within each cancer type.  \textbf{B.} A latent autoencoder (AE) and a latent diffusion model (LDM) are trained on a large-scale dataset of histopathology images paired with prototype values obtained from clustering for conditional image synthesis under the guidance of a trained latent classifier. \textbf{C.} Sampling a fixed number of images from the LDM, guided by each prototype, to construct a synthetic dataset for SSL model training. \textbf{D.} We test the proposed method and baselines with few-shot learning on clinical downstream tasks (subtyping and survival prediction).}
    \label{fig:figure1} 
\end{figure*}
Although diffusion models have shown promise in generating synthetic histopathology images, existing approaches cannot guarantee clinically meaningful diversity in their outputs. Our work addresses this limitation by incorporating structured prior knowledge to produce both diverse and realistic samples. We introduce a prototype-guided diffusion model to generate high-quality synthetic histopathology images, thus ensuring clinically meaningful diversity. Rather than assuming that every real image contains equally valuable information, we use a clustering technique to identify prototypical tissue types. These prototypes distill a large and complex dataset into its essential components, effectively capturing the core concepts needed to describe the underlying pathology. By incorporating these prototypes into training, we ensure that subsequent synthetic data remains both biologically meaningful and diverse. We then pre-train an SSL framework on the generated dataset and evaluate its effectiveness through multiple downstream pathology classification tasks, including subtyping and survival prediction, using an attention-based multiple instance learning (ABMIL) framework \cite{ilse2018attention}. Our results demonstrate that with a relatively tiny amount of real data, our techniques can produce downstream predictive models that have comparable performance to those that require massive amounts of real data. Our methodology may expedite model development by more efficiently utilizing small datasets, ultimately increasing the pace of innovation in the field of digital pathology. 

Our contributions are: 1) a novel synthetic data curation method for SSL in digital pathology, using diffusion models guided by histological prototypes derived from clustering; 2) the first digital pathology foundation model trained on 1.7 million synthetically generated images, achieving performance comparable to models trained on clinical datasets that are 60 times larger; 3) a comprehensive evaluation of five subtyping and three survival tasks across various cancer types.

\section{Related work}


\subsection{Data curation for self-supervised learning}
Previous efforts in data curation for SSL have emphasized the importance of constructing datasets that are extensive in size and encompass a diverse range of samples while maintaining a balanced distribution across disease categories \cite{vo2024automatic}. To achieve balanced data subsampling, a popular strategy is to use clustering techniques such as k-means, where centroids serve as representative points for data subsets. A hierarchical k-means clustering algorithm for automated balanced dataset curation from uncurated data was proposed by Vo et al. \cite{vo2024automatic}, yielding substantial downstream performance improvements when self-supervised learning features were trained on the resulting curated datasets.

Van et al. \cite{van2024graph} took a different approach and proposed solving a problem of balanced data subset selection as a graph-matching task where the goal is to select the most distinct subset utilizing pairwise similarities. The authors showed substantial downstream performance improvement by training a self-supervised learning algorithm on the rebalanced dataset. 

Given the critical role of data diversity and balance in SSL, our work extends these ideas by leveraging histological prototypes to guide the generation of synthetic pathology datasets. Unlike prior efforts that focus solely on selecting subsets from existing data, our approach actively generates new samples to enhance morphological diversity while preserving diagnostic relevance.

\subsection{Histological prototypes}
Tissue prototypes can be identified as clusters of morphologically similar regions within tissue samples \cite{claudio2024mapping, song2024morphological}. These prototypes are typically represented as centroids derived from clustering, capturing distinctive morphological features present in the tissue. Utilizing tissue prototypes has shown success in recent applications in digital pathology, e.g., prototypical MIL (DeepAttnMISL) \cite{yao2020whole}, which aggregates patch embeddings within the same cluster, followed by aggregating the pooled cluster embeddings. Recently, an unsupervised method that utilizes a Gaussian mixture model to create compact, interpretable representations of whole slide images through morphological prototypes was proposed by Song et al. \cite{song2024morphological} as an alternative to MIL approaches. Extensive evaluation demonstrated its competitive performance against supervised methods.

Building upon these prior works, our framework integrates histological prototypes into a generative model, ensuring that synthetic pathology images preserve biologically meaningful patterns. Unlike previous methods that use prototypes primarily for feature extraction, we incorporate them directly into the image generation process, creating a synthetic dataset that reflects real-world histological diversity.

\subsection{Conditional diffusion models for data augmentation}
While unconditional diffusion models can generate diverse images without explicit supervision, conditional diffusion models offer greater control over the generated outputs, making them more suitable for clinically meaningful applications \cite{dhariwal2021diffusion, zhan2024conditional}. For example, Yellapragada et al. proposed to use histopathology reports to condition diffusion model training and achieved a SOTA text-to-image generation performance \cite{yellapragada2024pathldm}.

\begin{figure}[]
    \centering
    \includegraphics[width=\linewidth]{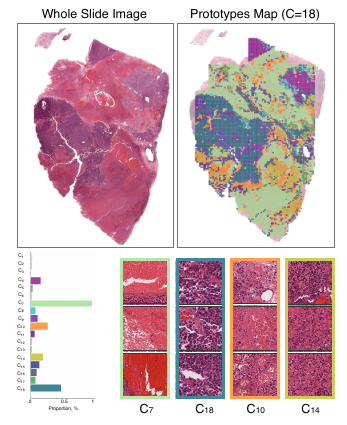} %
    \caption{Example of a WSI from TCGA-UCS (Uterine Carcinosarcoma) with its corresponding prototype map, showing 18 detected clusters for this cancer type, along with prototype distribution for the slide and patch examples from the four largest clusters.}
    \label{fig:figure2} 
\end{figure}

However, training conditional models requires a large amount of labeled data, which is often difficult to obtain in medical imaging domains. Ye et al. \cite{ye2023synthetic} addressed this challenge by introducing a two-stage diffusion-based method for high-quality digital pathology data generation, involving initial unconditional latent diffusion model training on a large unlabeled dataset, followed by fine-tuning on a smaller labeled cohort. The original dataset was augmented with data generated by the proposed method, which led to a significant 6.4\% classification improvement in one downstream task.

In this work, we propose an extension of this method by introducing an unsupervised histological prototype-guided approach to generate a comprehensive synthetic dataset that is used to pretrain an SSL model. Unlike the previous method that requires a small labeled subset for conditional generation, our approach uses unsupervised prototypes, enabling controlled data generation without the need for labeled examples.
This approach reduces reliance on clinical samples while maintaining strong downstream performance, offering a scalable solution for training robust pathology models with minimal real data.

\begin{figure*}[h!]
    \centering
    \includegraphics[scale=1]{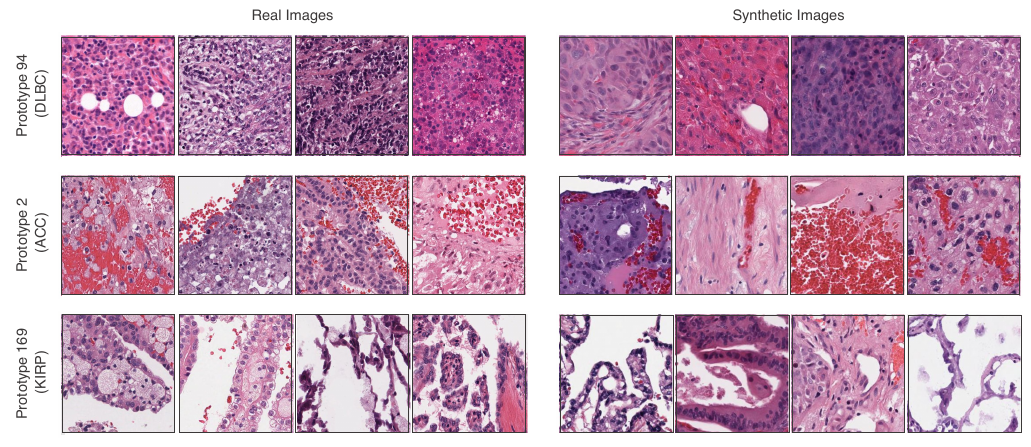} %
    \caption{Comparison of real images from the training subset with images generated using prototype guidance for three prototypes, each representing a different tissue type: DLBC (Diffuse Large B-Cell Lymphoma), ACC (Adrenocortical Carcinoma), and KIRP (Kidney Renal Papillary Cell Carcinoma).}
    \label{fig:figure3} 
\end{figure*}
\section{Methods}
In this study, we propose a prototype-guided diffusion framework for generating synthetic histopathology data, and we demonstrate its utility in SSL and downstream tasks (see Figure \ref{fig:figure1}). We first identify histological prototypes by clustering to find morphologically similar regions within tissue samples. Using the extracted prototypes, we then train a conditional latent diffusion model to generate high-fidelity synthetic pathology images. The generated dataset is used to pre-train an SSL framework, learning rich feature representations without requiring labeled pathology data. This framework enables the development of generalizable embeddings that capture essential histological features. Finally, we evaluate the effectiveness of the SSL-pretrained features by training an ABMIL model to aggregate patch-level features and make slide-level predictions.

\subsection{Histological prototypes}
Starting with a collection of WSIs $X_j, _{j=1, \dots, J}$  from the pretraining cohort $\mathcal{D}$ spanning $J$ organs, we extract non-overlapping patches \( X_j = \{ x_j^1, \dots, x_j^{N_j} \} \), where each patch \( x_j^n \in \mathbb{R}^{W \times H \times 3} \), and transform them into compressed feature embeddings $\{ h_j^1, \dots, h_j^{N_j} \}$ using an encoder pre-trained on $\mathcal{D}$ \cite{chen2022scaling}.

We utilize k-means clustering within each organ-specific subset $X_j$. Rather than clustering the entire subset directly, we first draw a uniform sample $\mathcal{D}^{'}_{i} \sim \mathcal{D}_{i}$, where $|\mathcal{D}^{'}_{i}| \ll |\mathcal{D}_{i}|$ and $1 \leq i \leq 32$. This sampling strategy mitigates computational inefficiency while preserving the representative structure of the data. We then apply fine-grained K-means clustering on each sampled subset:

\begin{equation} S_{i} \leftarrow \text{K-means} (\mathcal{D}^{'}_{i}) \end{equation}

where the final set of learned cluster centers is given by $S = \bigcup S_{i}$.

To identify the optimal number of clusters within each organ-specific cohort, we calculated the within-cluster sum of squares (WCSS). Using the "elbow" method, we determined the point at which further increases in the number of clusters led to WCSS diminishing. This approach enabled us to balance capturing meaningful biological variability while preventing unnecessary fragmentation of the data.

\subsection{Classifier-guided latent diffusion model}
We utilize Latent Diffusion Models (LDM) \cite{rombach2022high}, which offer a more computationally efficient alternative to previous approaches \cite{dhariwal2021diffusion} while maintaining performance quality. LDM use a two-stage process, beginning with training a latent autoencoder (AE) \cite{kingma2013auto} to compress images into a lower-dimensional latent space. Then, a diffusion model is trained to generate images by learning the distribution of these latent representations. 

Specifically, given an image \( x_0 \in \mathbb{R}^{H \times W \times 3} \), a pretrained latent autoencoder \( \mathcal{E} \) maps it to a latent representation \( z_0 \):

\begin{equation}
z_0 = E(x_0), \quad z_0 \in \mathbb{R}^{h \times w \times c},
\end{equation}

where \( h, w, \) and \( c \) are the dimensions of the latent space. The forward diffusion process gradually adds Gaussian noise to \( z_0 \) over \( T \) timesteps:

\begin{equation}
q(z_t | z_{t-1}) = \mathcal{N}(z_t; \sqrt{1 - \beta_t} z_{t-1}, \beta_tI),
\end{equation}

where \( \beta_t \) controls the noise scale at each step. The reverse process learns to reconstruct the latent representation by parameterizing:

\begin{equation}
p_\theta(z_{t-1} | z_t) = \mathcal{N}(z_{t-1}; \mu_\theta(z_t, t), \Sigma_\theta(z_t, t)),
\end{equation}

where \( \mu_\theta(z_t, t) \) and \( \Sigma_\theta(z_t, t) \) are the predicted mean and covariance.  

To introduce classifier guidance, we train a classifier \( C_\phi(y | z_t, t) \) on latent representations, which helps refine the generation process. The mean update in the reverse process is modified as:

\begin{equation}
\tilde{\mu}_\theta(z_t, t, y) = \mu_\theta(z_t, t) + w \cdot \Sigma_\theta(z_t, t) \nabla_{z_t} \log C_\phi(y | z_t, t),
\end{equation}

where \( w \) is the guidance scale that controls how strongly the classifier influences generation.  The reverse diffusion process generates a novel latent \( \tilde{z}_0 \) satisfying the class condition \( y \) through a Markov chain that starts from Gaussian noise \( z_T \sim \mathcal{N}(0, I) \), using the following class-conditioned sampling step:

\begin{equation}
p_{\theta, \phi}(z_{t-1} | z_t, y) = \mathcal{N}(z_{t-1}; \hat{\mu}_\theta(z_t | y), \Sigma_\theta(z_t)).
\end{equation}

Finally, after sampling a latent \( z_0 \), the decoder \( D \) reconstructs the image:

\begin{equation}
\hat{x}_0 = D(z_0).
\end{equation}

In our application, \( C_\phi \) is trained to classify histological prototypes, ensuring that the generated synthetic pathology images maintain biologically meaningful variations while enhancing data diversity.

\begin{table*}[]
\centering
\begin{tabular}{l|cc|cc|cc|cc|cc}
              & \multicolumn{2}{c|}{Lung}       & \multicolumn{2}{c|}{Prostate}   & \multicolumn{2}{c|}{Lymph Nodes} & \multicolumn{2}{c|}{Ovarian}    & \multicolumn{2}{c}{Breast}      \\ \hline
              & \multicolumn{2}{c|}{PLCO}       & \multicolumn{2}{c|}{PANDA}      & \multicolumn{2}{c|}{Camelyon}    & \multicolumn{2}{c|}{UBC-OCEAN}  & \multicolumn{2}{c}{PLCO}        \\ \hline
              & AUC            & F1             & AUC            & F1             & AUC             & F1             & AUC            & F1             & AUC            & F1             \\ \hline
UNI           & {\ul 0.971}    & 0.893          & \textbf{0.932} & {\ul 0.668}    & {\ul 0.960}     & 0.817          & \textbf{0.978} & \textbf{0.891} & \textbf{0.797} & {\ul 0.550}    \\
Prov-GigaPath & 0.970          & {\ul 0.901}    & 0.725          & 0.339          & 0.930           & \textbf{0.883} & 0.942          & 0.759          & 0.696          & 0.504          \\
CONCH         & 0.970          & \textbf{0.913} & 0.909          & 0.623          & 0.818           & 0.744          & 0.965          & 0.829          & 0.767          & 0.516          \\
iBOT-Synth    & 0.967          & 0.875          & 0.918          & 0.648          & 0.949           & 0.798          & 0.962          & 0.821          & 0.738          & 0.525          \\
iBOT-Hybrid   & \textbf{0.974} & 0.893          & {\ul 0.929}    & \textbf{0.673} & \textbf{0.965}  & {\ul 0.878}    & {\ul 0.972}    & {\ul 0.833}    & {\ul 0.785}    & \textbf{0.583} \\ \hline
\end{tabular}
\caption{\textbf{Subtyping prediction} results of iBOT-Synth (fully synthetic dataset) iBOT-Hybrid (synthetic + TCGA)  and baselines for five different subtyping tasks.  
The best performance is in bold, and the second best is underlined. 
}
\label{table1}
\end{table*}

\subsection{Slide encoding}
For a given histology slide, we adopt the ABMIL training paradigm \cite{ilse2018attention, li2021multi}, which involves partitioning the slide into smaller patches, extracting feature representations for each patch using a pre-trained vision encoder, and subsequently aggregating these patch-level embeddings to generate a holistic slide-level representation using a trainable attention-based pooling mechanism.

\textbf{Pre-trained feature encoder:}
We trained from scratch a ViT-Base (86 million parameters)
with iBOT \cite{zhou2021ibot} on 1.4 million synthetically generated H\&E patches sampled uniformly from each tissue prototype. So far, this is the largest SSL model trained solely on synthetically generated data. We denote feature embeddings for slide $X_{i}$ as $H_{i} \in \mathbb{R}^{N_{H} \times d_{h}}$, where $N_{H}$ - number of patches and $d_{h}$ - dimensionality of each patch-based feature vector.

\textbf{ABMIL slide encoding:}
We employ the widely used ABMIL model, which learns patch-specific attention weights to selectively aggregate patch embeddings $H_{i}$ into a comprehensive slide representation $h_{i}$.

\section{Experiments and Results}
\subsection{Datasets}
For prototype extraction and LDM training, we utilized FFPE (formalin-fixed, paraffin-embedded) H\&E-stained diagnostic slides from 32 cancer types within The Cancer Genome Atlas (TCGA) dataset, leveraging the consistently processed and patched data released as part of the CPIA dataset \cite{ying2023cpia}. The patches were originally 384×384×3, but a center crop was applied to resize them to 224×224×3.

The prototype extraction process yielded a total of 578 prototypes spanning 32 distinct cancer types (see example in Figure \ref{fig:figure2}).
We trained two separate versions of iBOT using two distinct datasets. First, a fully synthetic dataset was created by sampling 3,000 patches per prototype using our prototype-guided LDM, resulting in 1,734,000 patches (iBOT-Synth). Second, a combined dataset was formed by augmenting the synthetic data with 3,000 randomly sampled patches per prototype from the TCGA dataset, resulting in 3,468,000 (iBOT-Hybrid).

We compared the performance of iBOT-Synth and iBOT-Hybrid to UNI using five downstream subtyping tasks, two survival, and one prostate cancer biochemical recurrence (BCR) tasks. Our six different subtyping tasks are Non-Small Cell Lung Carcinoma (NSCLC) subtyping on
PLCO \cite{zhu2013prostate} (two classes), ISUP grading based on Prostate cancer grade assessment (PANDA) challenge \cite{bulten2022artificial} (six
classes), classification of breast cancer metastases in lymph nodes (Camelyon16) \cite{bejnordi2017diagnostic, litjens20181399} (three classes), ovarian cancer subtypes classification (UBC-OCEAN) challenge \cite{farahani2022deep, asadi2024machine} (five classes), and PLCO Breast cancer subtyping classification (three classes) \cite{zhu2013prostate}. Survival tasks included Breast Invasive Carcinoma (BRCA) and Non-Small Cell Lung Carcinoma (NSCLC) from PLCO. One private dataset from our institution was used for prostate cancer BCR prediction. For all downstream tasks, we performed a patient-level, label-stratified split into training, validation, and test sets with a 70:10:20 ratio unless a predefined split was provided.

\subsection{Implementation Details}
We utilized publicly available implementations and pre-trained weights for all baseline foundation models used for comparison,  following the official preprocessing pipelines provided in their respective repositories to maintain consistency with their original training procedures. No further fine-tuning or additional training was applied beyond the publicly released checkpoints.

For the downstream tasks, ABMIL models were trained from scratch using an identical set of hyperparameters across all downstream experiments. We employ a weight decay of $1 \times 10^{-5}$ and use the AdamW optimizer with a learning rate of $1 \times 10^{-4}$, along with a cosine decay scheduler. For the slide classification experiments, we utilized a cross-entropy loss. We employed early stopping if the validation loss failed to improve over ten consecutive epochs with total training epochs of 20. For survival prediction experiments, we used negative log-likelihood loss (NLL). ABMIL architecture used in the downstream experiments consists of three components. First, a 2-layer MLP with 256 or 512 hidden units, layer normalization, ReLu activation, and 0.25 dropout. This is followed by a gated-attention network consisting of 2-layer
MLP, with Sigmoid and Tanh activation, respectively, and 0.25 dropout. Finally, a post-attention linear classification layer with 256 or 512 hidden units is applied.

\subsection{Results}

We used the Fréchet Inception Distance (FID) score \cite{heusel2017gans} to assess the image quality of the synthetic samples, with the achieved value of 0.12 indicating high image quality \cite{ye2023synthetic}. Additionally, we qualitatively evaluated the results of the prototype-guided diffusion model used to generate data for iBOT-Synth and iBOT-Hybrid (see Figure \ref{fig:figure3}).

We evaluated our iBOT-Synth and iBOT-Hybrid models against three baseline encoders, noting significant dataset size variations: UNI \cite{chen2024towards} (100 million images, $\thicksim$60x larger than iBOT-Synth), CONCH \cite{lu2024visual} (1.17 million image-caption pairs, comparable to iBOT-Synth), and Prov-GigaPath \cite{xu2024whole} (1.3 billion images,  $\thicksim$760x larger than iBOT-Synth). Features extracted from these encoders were used within an ABMIL framework trained to perform cancer subtyping and survival prediction.
To assess the statistical significance between subtyping models, we perform a Wilcoxon Signed-Rank Test to evaluate the observed differences in performance between the two models \cite{xu2024whole}. To statistically compare the survival models, we use DeLong’s test to compare the concordance index (c-index) \cite{delong1988comparing}.

\subsubsection{Cancer Subtyping}
Both iBOT-Synth and iBOT-Hybrid models show competitive performance across all cancer types in both AUROC and F1-score metrics (see Table \ref{table1}). Notably, iBOT-Hybrid trained on $\thicksim$30 times fewer data points than UNI and $\thicksim$380 times fewer data points than Prov-GigaPath outperforms both models in AUC in lung and lymph node cancer subtyping tasks and shows the second highest AUC among all other cancer types. iBOT-Hybrid results in the highest F1 for prostate and breast subtyping tasks and shows the second highest F1 for lymph nodes and ovarian tasks. Compared to UNI, iBOT-Hybrid demonstrates significantly better performance in the lung ($\emph p$=0.016) and lymph nodes ($\emph p$$<$0.01)
cancer subtyping tasks. Additionally, there is no significant difference between iBOT-Hybrid and UNI for the ovarian ($\emph p$=0.14) cancer subtyping task. Statistical testing shows no significant difference between iBOT-Hybrid and CONCH for lung ($\emph p$=0.11) and ovarian ($\emph p$=0.053) cancer subtyping tasks. iBOT-Hybrid significantly outperforms CONCH for prostate ($\emph p$$<$0.01), lymph nodes ($\emph p$$<$0.01), and breast ($\emph p$$<$0.01) cancer subtyping tasks. Compared to Prov-GigaPath, iBOT-Hybrid demonstrated no significant performance difference in the lung cancer subtyping task ($\emph p$=0.24) but achieved significantly better results in all other subtyping tasks.

Statistical significance testing confirmed that while iBOT-Synth, trained entirely on synthetic data, was slightly outperformed by iBOT-Hybrid, it still achieved competitive performance. Notably, UNI was trained on $\thicksim$60 times more data, while Prov-GigaPath utilized $\thicksim$760 times more data. CONCH, trained on approximately the same amount of data, incorporated an additional modality, introducing an extra challenge compared to iBOT-Synth. Despite these significant differences in training data, there was no statistically significant difference between the prediction of iBOT-Synth and UNI on three out of five cancer subtyping tasks: lung ($\emph p$=0.25), prostate ($\emph p$=0.36), and ovarian ($\emph p$=0.22). iBOT-Synth showed no significant difference in performance compared to CONCH for three out of five cancer subtyping tasks: lung ($\emph p$=0.77), breast ($\emph p$=0.23), and ovarian ($\emph p$=0.21). iBOT-Synth did show significantly different performance than CONCH on the prostate ($\emph p$$<$0.01) and lymph nodes ($\emph p$$<$0.01) tasks. Compared to Prov-GigaPath, iBOT-Synth demonstrated no significant performance difference in the lung cancer subtyping task ($\emph p$=0.58) but achieved significantly better results in prostate classification ($\emph p$$<$0.01), lymph nodes ($\emph p$$<$0.01), ovarian ($\emph p$$<$0.01), and breast ($\emph p$$<$0.01) cancer subtyping tasks.


\begin{table}[h!]
\centering
\begin{tabular}{l|c|c|c}
              & Lung    & Prostate BCR & Breast  \\ \hline
              & PLCO    & Private      & PLCO    \\ \hline
              & c-index & c-index      & c-index \\ \hline
UNI           & \underline{0.578}   & 0.632        & 0.560   \\
Prov-GigaPath & 0.595   & 0.664        & 0.560   \\
CONCH         & 0.544   & 0.603        & 0.555   \\
iBOT-Synth    & 0.572   & \underline{0.7}          & \textbf{0.585}   \\
iBOT-Hybrid   & \textbf{0.636}   & \textbf{0.704}        & \underline{0.580}   \\ \hline
\end{tabular}
\caption{\textbf{Survival and prostate BCR prediction} results of iBOT-Synth (fully synthetic dataset) iBOT-Hybrid (synthetic + TCGA)  and baselines for two survival and one prostate BCR task.  
The best performance is in bold, and the second best is underlined. 
}
\label{table2}
\end{table}

\subsubsection{Survival Prediction and Prostate BCR Analysis}
We used the c-index for survival and prostate cancer BCR evaluation (see Table \ref{table2}).
In lung cancer survival prediction, iBOT-Hybrid achieves the highest c-index of 0.636, indicating the highest performance among all models. Prov-GigaPath also demonstrates strong performance with a c-index of 0.595, while UNI achieves 0.578. iBOT-Synth shows a competitive result of 0.572, suggesting that even with fully synthetic data, it maintains reasonable predictive power. CONCH, however, shows significantly lower performance in this task. iBOT-Hybrid significantly outperformed UNI ($\emph p$$<$0.01), Prov-GigaPath ($\emph p$$<$0.01) and CONCH ($\emph p$$<$0.01). iBOT-Synth significantly outperformed CONCH ($\emph p$$<$0.01) but showed no significant difference when compared to UNI ($\emph p$=0.12) and Prov-GigaPath ($\emph p$=0.32).

For prostate BCR prediction, both iBOT-Synth and iBOT-Hybrid achieve the highest c-index values, with 0.70 and 0.704, respectively. This indicates that our models are particularly effective in predicting prostate BCR, demonstrating statistically significant improvements over the three baseline methods. Prov-GigaPath shows a c-index of 0.664, while UNI achieves 0.632. CONCH again shows lower performance. 

In breast cancer survival prediction, iBOT-Synth achieves the highest c-index of 0.585. iBOT-Hybrid also performs well with a c-index of 0.580. UNI and Prov-GigaPath both achieve 0.560, while CONCH shows 0.555. iBOT-Hybrid didn't show a statistically significant difference with UNI ($\emph p$=0.055) and Prov-GigaPath ($\emph p$=0.061), but significantly outperformed CONCH ($\emph p$$<$0.01). iBOT-Synth significantly outperformed UNI ($\emph p$$<$0.01), Prov-GigaPath ($\emph p$$<$0.01) and CONCH ($\emph p$$<$0.01).

\section{Conclusions}
In conclusion, this study explores the potential of a prototype-guided diffusion model to generate high-fidelity synthetic pathology data, reducing the need for large, real-world datasets in digital pathology. Our approach enables large-scale self-supervised learning while ensuring biologically meaningful variations in the generated data. We demonstrate that self-supervised features trained on our synthetic dataset achieve competitive performance, using up to 760 times less data than models trained on large real-world datasets. Notably, our models showed statistically comparable or superior performance across various evaluation metrics. Additionally, combining synthetic and real data further enhanced performance, achieving top results in several downstream tasks. These findings highlight the effectiveness of generative AI in reducing reliance on extensive clinical datasets, offering a more efficient approach for training models in digital pathology.
{
    \small
    \bibliographystyle{ieeenat_fullname}
    \bibliography{main}
}


\end{document}